\documentclass[twocolumn,prb,showpacs,preprintnumbers,amsmath,amssymb]{revtex4}

\usepackage{graphicx}
\usepackage{dcolumn}
\usepackage{bm}
\usepackage{multirow}

\begin{document}
\preprint{Journal of Physics: Condensed Matter}
\title{Ultrafast Carrier Dynamics in Ag-CdTe Hybrid Nanostructure: Non-radiative and Radiative Relaxations}
\author{Sabina Gurung$^{ab}$}
\email{sabusg12@gmail.com}
\author{Durga Prasad Khatua$^{ab}$}
\author{Asha Singh$^{a}$}%
\author{J. Jayabalan$^{ab}$}
\affiliation{$^{a}$ Nano Science Laboratory, Materials Science Section, 
	Raja Ramanna Centre for Advanced Technology, Indore, India - 452013. \\
		$^{b}$ Homi Bhabha National Institute, Training School Complex, Anushakti Nagar, Mumbai, India - 400094.}
\date{\today}

\begin{abstract}
In this article, we study non-radiative and radiative relaxation processes in a hybrid formed by combining Ag nanoparticle (NP) and CdTe quantum dots (QD) using transient transmission spectroscopy. The ultrafast transient transmission of hybrid, when excited at 400 nm, shows a faster recovery of hot electrons at a shorter time scale (few picoseconds) while it shows a slower recovery at longer time scale (few tens of picoseconds). Further it is found that the contribution of CdTe QD to the transient transmission is increased in the presence of Ag NP. However, the radiative relaxation in CdTe QDs get quenched in the presence of Ag NP. This work provides significant insight into the various relaxation processes that leads to the charge transport and PL quenching mechanisms in metal-semiconductor hybrids.
\end{abstract}

\pacs{Valid PACS appear here}
\keywords{Hot-electrons \and Transient \and Metal Nanoparticles \and Semiconductor Quantum dots \and Hybrid \and Electron-Phonon coupling}
\maketitle

\section{Introduction}
Metal-semiconductor hybrid nanostructure (HNS) shows unique properties that can supersede the combined function of individual material because of its synergistic behavior. A metal nanoparticle (NP) when excited at its localized surface plasmon resonance (LSPR), can modify the local field distribution around it. The light-matter interaction in a semiconductor quantum dots (QD) can be enhanced by placing it near to the metal NP. Such enhancement can increase the efficiency of semiconductor devices like photodetector, solar cells, LED etc\cite{Plasmon-enhanced-detector-2018,Plasmon-enhanced-devices,Hybrid_plasmon_spectroscopy_Jiang_AdvanceMat_2014}. It is also possible that charges excited in the metal NP can move to the semiconductor QD in ultrafast time scale\cite{Samanta-ultrafast-AgCdTe-JPCC-2016,Counting_electron_Jayabalan_2019}. Such charge transfer can further increase the efficiency of light harvesting devices\cite{LED_MRSbulletin_2013,Wu632}. Conversely, metal NP can act as an electron sink that can disturb the electron-hole recombination process in the semiconductor QD resulting in quenching of PL\cite{Au-electron-sink-2013-AppliedCatalysisB,Electron-sink-2013-JMCA}. The carrier dynamics that takes place between metal NP and semiconductor QD plays a vital role in determining the final optical response of HNS. The ultrafast response of hybrid can have both the plasmonic and excitonic contributions. Both of these contributions have different physical origin and are usually described by different physical theories. In the case of isolated metal NPs, ultrafast optical response can be explained by the increase in temperature of free electrons\cite{Jayabalan-2011-JOSAB}. On the other hand, in the case of isolated semiconductor QDs, it is explained by band to band absorption, band filling, inter and intra band relaxation, etc \cite{Ultrafast-Absorption-II-VI-Nanowires-2009-JPCC}. In semiconductor QDs, both radiative and non-radiative processes contribute to the total carrier dynamics, while in metal NP only non-radiative processes contribute. Thus the various dynamical processes that occurs in individual materials in HNS as well as their synergistic interaction complicates the understanding of its complete optical response. Nonetheless, it is important to comprehend the basic interactions between these individual constituents for designing an HNS focused toward a particular application. 

In this article, we study the origin of the ultrafast optical response of metal-semiconductor HNS using a colloidal mixture of Ag NP and CdTe QD, labeled as Ag-nCdTe, using ultrafast transient spectroscopy. Ag-nCdTe HNS formed by Ag NPs and CdTe QDs is an excellent candidate to study the optical response under a strong exciton-plasmon coupling regime because the plasmon band of Ag NP is well separated from the band edge emission of CdTe QD. Besides, CdTe QDs has a fast electron injection time constant compared to other II-VI semiconductor like CdSe\cite{CdSe-CdTe-ACSnano-2009}. To reveal the origin of optical process in HNS, we excite the carriers in Ag-nCdTe hybrid system and probe the carrier dynamics near Ag NP plasmon peak and exciton peak in CdTe QD. We study the changes in optical properties with respect to the plasmonic response by keeping the concentration of CdTe QDs in colloid low enough such that we do not get any measurable response from the bare CdTe QDs colloid. The ultrafast dynamics of Ag-nCdTe when probed at plasmon peak of Ag NP and deep conduction band in CdTe QD is different from that when probed near to the band edge of the CdTe QD and away from the plasmon band. We have also studied the PL spectra and time-resolved PL of Ag-nCdTe hybrid colloid and compared it with that of bare CdTe QD colloid. Lastly the origin of the observed optical response is explained using electromagnetic theory, two temperature model and hot-electron transport process.

\section{Results and Discussions}
\begin{figure}
	\includegraphics[width=0.5\textwidth]{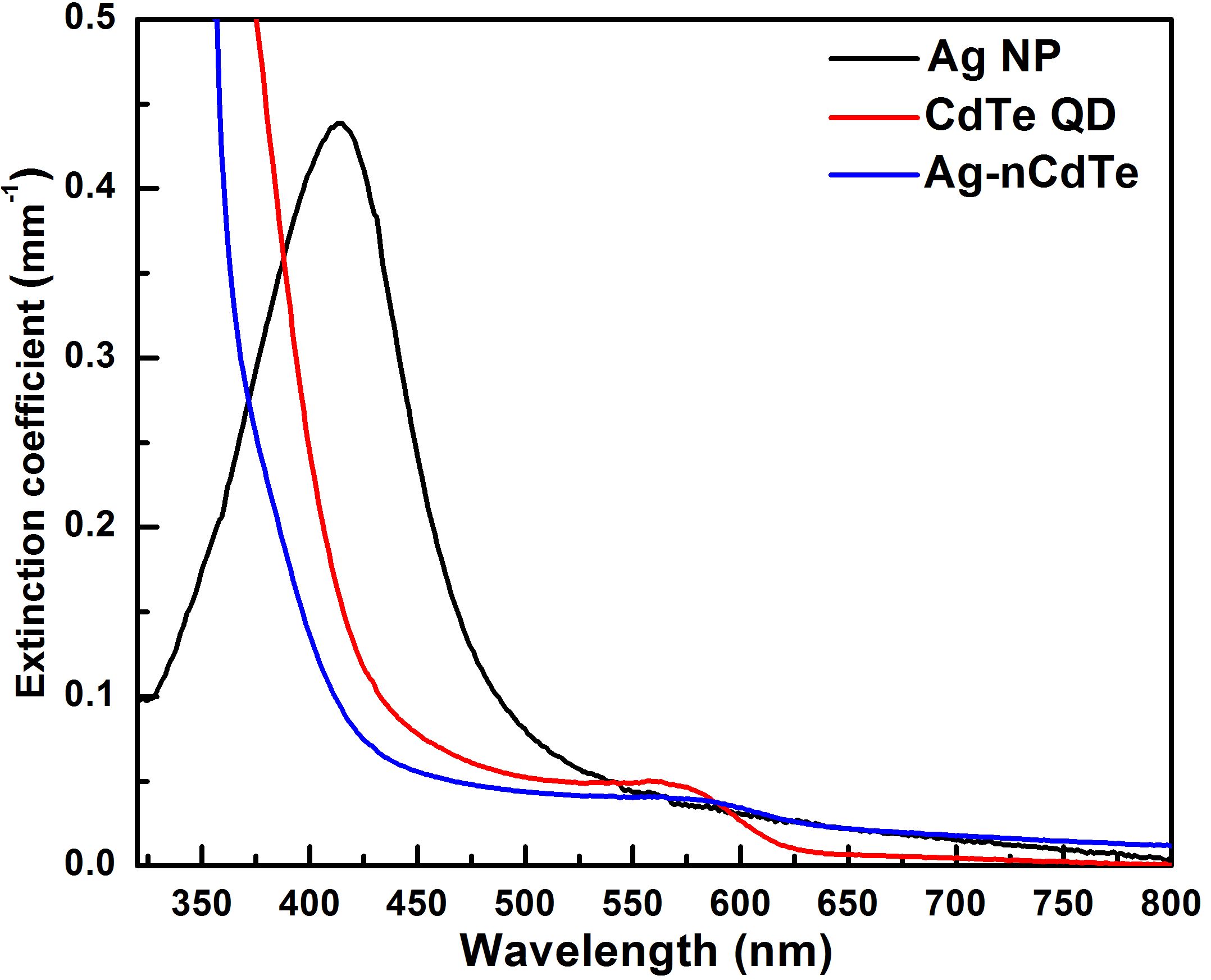}
	\caption{Extinction spectra of the colloidal solutions of Ag NP, CdTe QD and Ag-nCdTe hybrid.}
	\label{ExtincCoeff}       
\end{figure}

The individual colloidal solutions of Ag NPs and CdTe QDs dispersed in water were prepared and characterized separately. The volume fraction of CdTe QDs and Ag NPs in water in the individual colloids were of the order of $10^{-6}$ and  $10^{-7}$, respectively. Fig.\ref{ExtincCoeff} shows the extinction spectra of Ag NP and CdTe QD colloids. The extinction spectrum of Ag NP colloid shows a well-known single LSPR peak at 413 nm, which is typical for small Ag spherical particles in water\cite{Absorption_Scattering_small_particles_borhen_1983}. The average size of Ag NPs corresponding to the LSPR is found to be 17 nm using Atomic Force Microscopy (AFM) topography measurement. The extinction spectrum of CdTe QD colloid shows the lowest energy excitonic peak, the 1s-1s transition, at 560 nm. The average size of the CdTe QDs estimated using Peng's formula is $\sim$ 3.4 nm and also confirmed by AFM measurement \cite{Pengs_cdte_quantum_dots_size_2003}. In the final colloidal solution, CdTe QDs are capped with thiol-glycolic acid (TGA) and Ag NPs are capped by trisodium citrate. Due to the low volume fractions and in presence of capping agents, the individual particles are expected to be well separated in their respective colloidal solutions. 

A mixing ratio of $\gamma$ = 0.56, where $\gamma$ is defined as the ratio of volume of as-prepared Ag NP colloid to that of CdTe QD colloid ($\gamma$ = V$_{P}$/V$_{D}$) was used to prepare HNS colloidal sample.
Although the capping agents TGA and trisodium citrate develops negative charges on CdTe QD and Ag NP; once mixed, the CdTe QDs get attached to the Ag NPs due to the higher affinity of TGA to the metal surface\cite{Ligand_exciton_photovoltaic_PbS_Jin_PCCP_2017, TGA_CdTe_Europium_surface_coordinated_emission_Gallagher_inorganic_chemistry_2013}. Thus, self-assembled Ag-nCdTe hybrid nanostructures of negatively charged CdTe QDs and Ag NPs are expected to form in the mixed colloid. Based on our previous study, at the mixing ratio of $\gamma$ = 0.56, in the Ag-nCdTe hybrid, each of the Ag NP is completely surrounded by 32 CdTe QDs\cite{Sabina-JAP-2018}.

Once these CdTe QDs surrounds the Ag NP surface, further attachment of CdTe QDs to the same Ag NP is prevented thus making a stable HNS in the mixed solution. Ideally in the Ag-nCdTe HNS sample it is expected that there are only HNS with no individual particles. Fig.\ref{ExtincCoeff} also shows the measured extinction spectrum of Ag-nCdTe hybrid colloids and is different from that of both of its constituent colloids\cite{Sabina-JAP-2018}. The Ag-nCdTe hybrid has an increasing  absorption strength towards UV regime similar to that of a bare CdTe QD colloid. Additionally, the Ag-nCdTe hybrid has an increased absorption in the longer wavelength regime (starting from $\sim$ 600 nm). The excitonic peak in case of Ag-nCdTe colloid also got bleached compared to the bare CdTe QD colloid. Various groups have reported similar changes in the extinction spectrum in metal-semiconductor HNS when compared to that of individuals\cite{Ag_cdte_selforganized_electrostatic_interaction_Wang_Spect_Acta_2005,Hybrid_Optoelectronics_Nahar_ACSNano_2015,Luminescent_CdTe_QD_2003,XIA2008166, Enhanced_PL_CdTe_Ag}. The increase in the absorption cross-section in the longer wavelength of Ag-nCdTe is attributed to the defect states created by the attachment of Ag NP to CdTe QD. On the other hand, the presence of MNP also dissociates the excitons in the semiconductor QD resulting in the reduction of exciton peak strength.

\begin{figure}
	\includegraphics[width=0.5\textwidth]{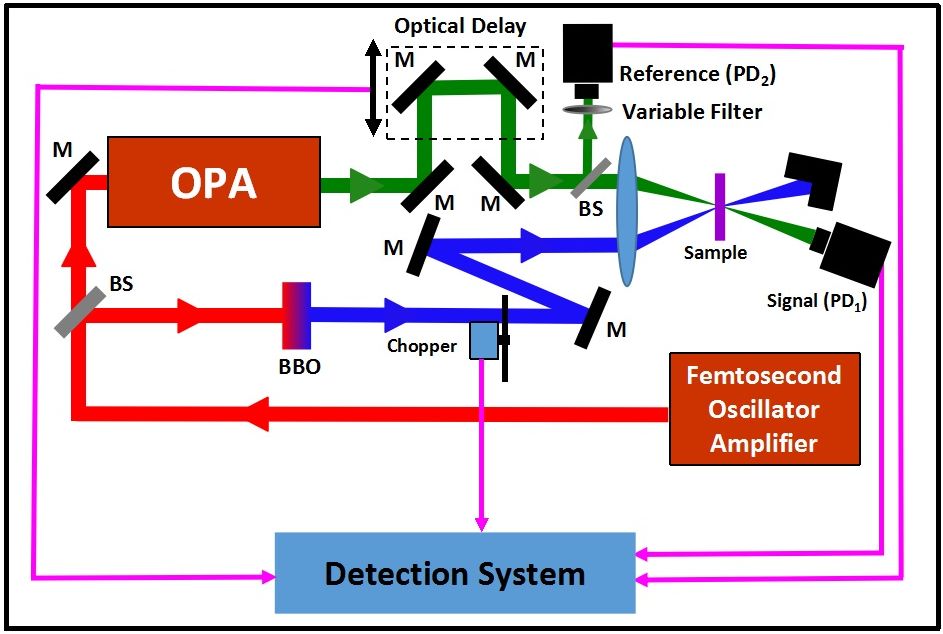}
	\caption{Schematic of the pump-probe setup used for the measurement of transient transmission from the samples.}
	\label{Setup}       
\end{figure}

To study the ultrafast transient phenomena in Ag-nCdTe hybrid nanostructure, transient transmission measurement were carried out in two-color pump-probe geometry (Fig.\ref{Setup}) using a 1 kHz Ti-Sapphire laser of pulse duration $\sim$ 35 fs operated at 800 nm wavelength\cite{Durga2020Filter}. In this setup, the output of the femtosecond oscillator-amplifier was split into two beams using a beam splitter. One of these beams was converted to 400 nm by second-harmonic generation in a Beta-barium Borate crystal and was used as the pump beam. The other beam from the beam splitter was fed to an optical parametric amplifier (OPA). The OPA output, centered either at 408 nm or 550 nm, was used as the probe beam. The wavelength of the pump beam was chosen such that it excites the sample near the LSPR of the Ag NP in water. Similarly, the probe was chosen such that it can measure the changes in the optical properties near the LSPR of Ag NP or near the band edge of the CdTe QD. The pump fluence at the sample location was 2.1 $\mu$Jmm$^{-2}$. Throughout the measurement, the colloidal samples were circulated in a 1 mm flow cell to avoid any thermal damage.

\begin{figure}
  \includegraphics[width=0.5\textwidth]{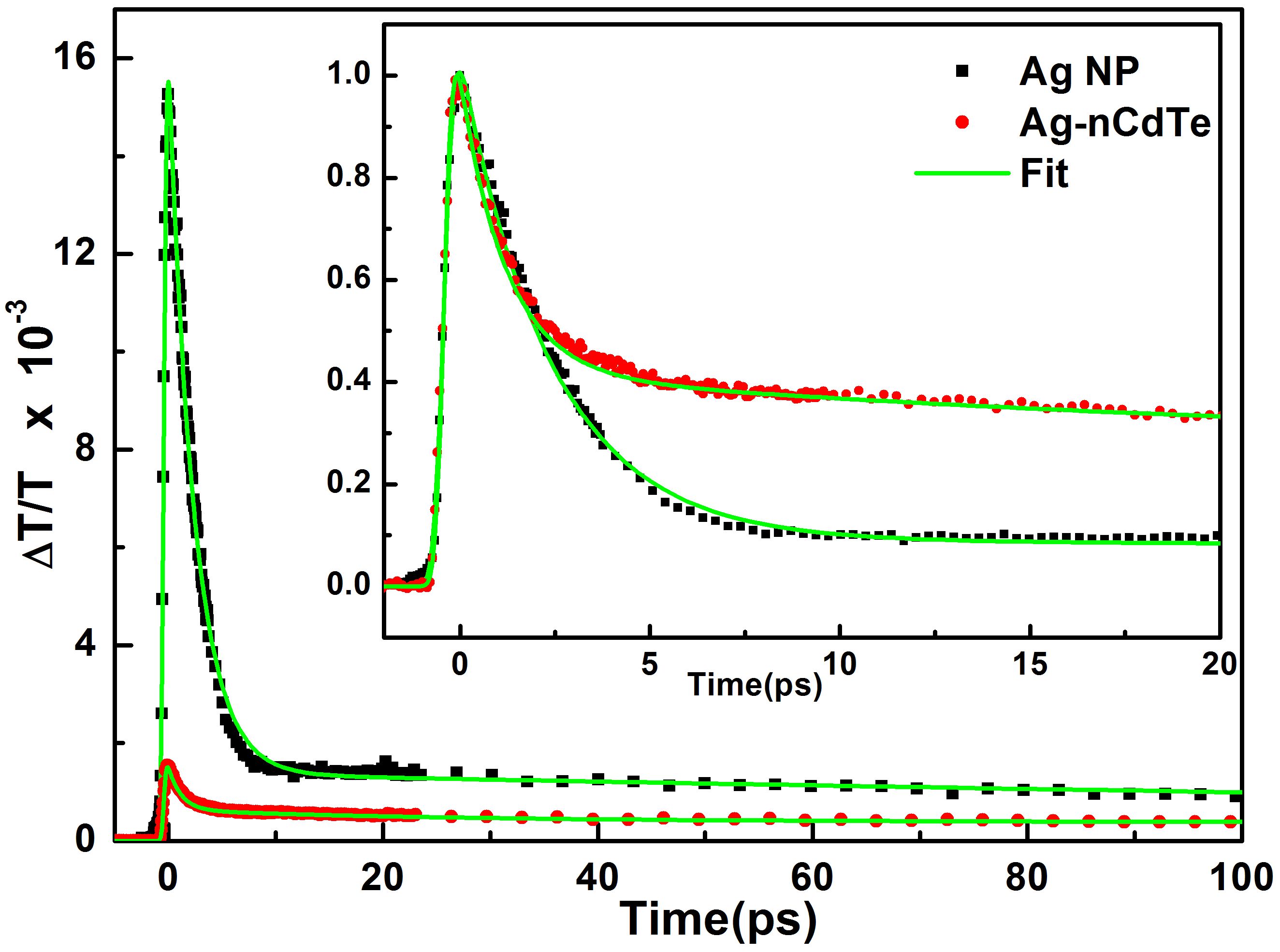}
\caption{Transient transmission of Ag NPs and Ag-nCdTe hybrid colloidal samples when excited at 400 nm and probe at 408 nm. Inset shows the normalized curve of the same data.}
\label{Fig:400-408}       
\end{figure}

Fig.\ref{Fig:400-408} shows the transient transmission ($\Delta T/T$) measured for the colloidal samples of bare Ag NP and Ag-nCdTe hybrid colloid when probed at 408 nm. Both Ag NP and Ag-nCdTe hybrid colloids showed measurable change in the transmission of the sample at intensity used in the measurement. The magnitude of $\Delta T/T$ in both the colloids start increasing with the arrival of the pump pulse, reaching a maximum by about 450 fs. At later time delay, $\Delta T/T$ reduces, leaving a residue even up to 100 ps. However, the peak of $\Delta T/T$ of Ag-nCdTe colloid when probed at 408 nm is only 10\% of that of Ag NP colloid. Further, the reduction in the amplitude of fast component is much higher than that of the longer decay component. This difference in the reduction in the magnitudes can be clearly seen in the normalized plot of the $\Delta T/T$ (inset of Fig.\ref{Fig:400-408}). Relatively, the long decay time component is much more in the case of Ag-nCdTe hybrid compared to that of bare Ag NP colloid. For the number density of CdTe QD same as that in the Ag-nCdTe hybrid, CdTe QD colloid did not show any measurable $\Delta T/T$ signal. Thus, the finite $\Delta T/T$ measured for the Ag-nCdTe hybrid should be attributed to the contribution by the Ag NP and the synergistic interaction between Ag NP and CdTe QDs.

Let us first look at the origin of transient transmission signal from Ag NP colloid. Over the decades, the origin of ultrafast optical response of the bare metal NP colloid has been studied by several groups\cite{Landua-damping-guillon-2004-ultrafast,Counting_electron_Jayabalan_2019,Non-equilibrium-electron-dynamics-nobel-PRB-2000,Electron-dynamics-Chemical-Physics-2000,Higher-nonlinearity-JOSAB-Jayabalan_2011}. When an ultrashort pulse excites an Ag NP at or near its LSPR, the absorbed energy creates a non-thermal energy distribution among the electrons which changes to thermal distribution in about 300 fs to 500 fs\cite{Non-equilibrium-electron-dynamics-nobel-PRB-2000,Landua-damping-guillon-2004-ultrafast}. At the end of the thermalization, the temperature of the free electrons is much higher than that of lattice. As the temperature of the free electrons increases, the real part of the dielectric constant of metal also increases\cite{Higher-nonlinearity-JOSAB-Jayabalan_2011}. It is well known that the peak of LSPR of metal nanosphere occur at a wavelength where $\varepsilon_m^r$ = - 2 $\varepsilon_h$. Here, $\varepsilon_m^r$ is the real part of dielectric constant of the metal and $\varepsilon_h$ is the dielectric constant of the host medium (water in the present case). Once hit by an ultrashort pulse, an increase in temperature of electrons will cause the LSPR peak to shift towards a longer wavelength causing a change in the transmission of the Ag NP colloid \cite{Higher-nonlinearity-JOSAB-Jayabalan_2011,Non-equilibrium-electron-dynamics-nobel-PRB-2000,Electron-dynamics-Chemical-Physics-2000}. Therefore the maximum in $\Delta T/T$ occur after the thermalization of electrons which is about $\sim$ 450 fs after the excitation by the pump pulse for the present Ag NP colloid. Further, when probed on the blue side (408 nm), a red-shift of LSPR will cause an increase in transmission (positive $\Delta T/T$). At later times the hot electrons heat the lattice through electron-phonon (e-ph) interaction, together reaching a much lower temperature than the initial hot electrons\cite{Electron-dynamics-Chemical-Physics-2000}. The cooling of electron can be observed as a nearly exponential decay of the magnitude of $\Delta T/T$ in picosecond timescales \cite{Electron-dynamics-Chemical-Physics-2000,Higher-nonlinearity-JOSAB-Jayabalan_2011}. An exponential fit to the measured data in the case of Ag NP colloid yields the e-ph thermalization time to be 2.3 ps at the pump fluence of 2.1 $\mu$Jmm$^{-2}$. At the end of e-ph thermalization, the temperature of the particle is higher than that of the surrounding host. At later times, the particle cools by delivering heat to the host which takes few hundreds of picoseconds\cite{Hot-electron-relaxation-JPCB-2002,Heat-transfer-medium-ChemPhys-2005}. This leaves a small residual, $\Delta T/T$, even after 100 ps delay between the pump and probe pulses.

Let us now look at the origin of the ultrafast optical response of the Ag-nCdTe hybrid colloid. In a metal-semiconductor hybrid, when a metal NP interacts with semiconductor QD, the optical response of the combined nanostructure is expected to be very different from that of the individual components. When excited at the LSPR, the local field near the metal NP can be much higher than the applied; such a large field can increase the absorption in semiconductor QD\cite{Absorption_enhancement_NanoLett_2012,Luo2019}. It is also possible that excited carriers may get transported from metal NP to semiconductor QD or vice versa\cite{Counting_electron_Jayabalan_2019}. Additionally, energy exchange processes like foster resonance energy transfer (FRET) and plasmon-induced resonance energy transfer (PIRET) are also possible in the hybrid system\cite{Energy_transfer_Qd_Au_2004,Controlling_PIRET_Metal_TiO2_JPCC_2015}. Coupling between the metal NP and semiconductor QD is also known to introduce additional defect states in the hybrid nanostructures\cite{Hybrid_Optoelectronics_Nahar_ACSNano_2015,Luminescent_CdTe_QD_2003}. The individual contribution from each of these above mentioned processes to the optical response of the hybrid depends on the shape, size and material of the individual components, the properties of linking molecules and their spatial distribution. To understand the origin of the optical response of the hybrid arising due to their synergistic interaction, it is necessary to understand changes in the LSPR of Ag NP due to the presence of CdTe QDs.

\begin{figure}
	\includegraphics[width=0.5\textwidth]{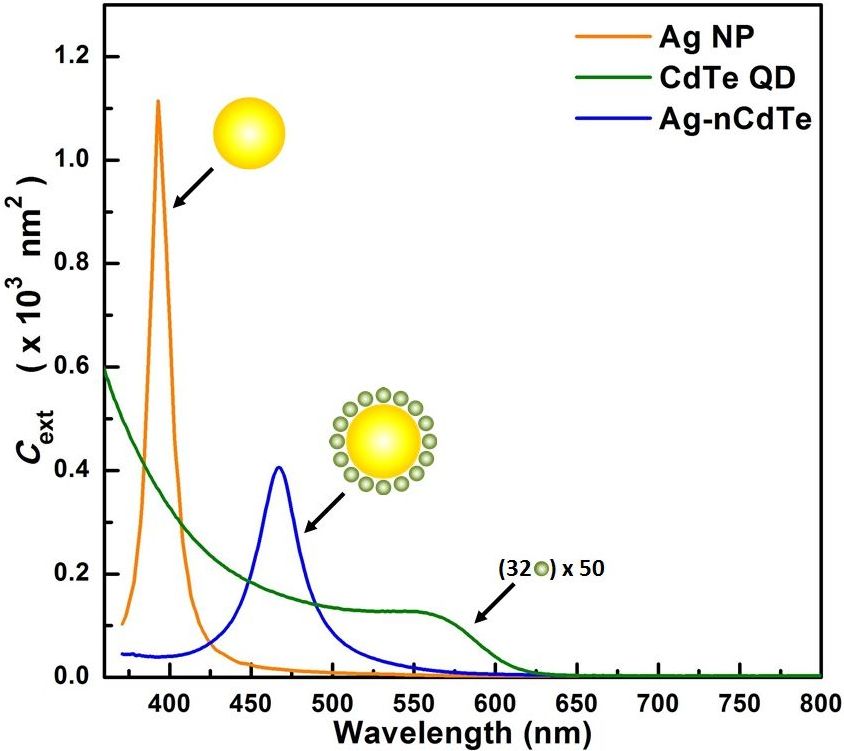}
	\caption{The extinction cross-section calculated using T-Matrix technique for a single Ag NP in water (orange), 32 non-interacting CdTe QDs and the hybrid nanostructure Ag-nCdTe (1 Ag surrounded in all directions by 32 CdTe QDs).}
	\label{Fig:TmatrixAgnCdTe}       
\end{figure}

It is well known that the LSPR response of a metal NP depends strongly on the dielectric constant of the surrounding medium. To understand the effect of the presence of CdTe QD on the optical response of Ag NP, we have calculated the optical response of the combined hybrid nanostructure using the T-matrix technique. T-matrix is a numerical technique to calculate the optical response of a collection of small spherical particles distributed in space using the dielectric constant of the individual particles. It should be noted that the optical response calculated using the T-matrix technique considers only electromagnetic interaction between the Ag NP and CdTe QD. However in a real metal-semiconductor hybrid, there is not only electromagnetic interaction but several other charge and energy exchange processes mentioned earlier. Thus a calculation which assumes only the electromagnetic interaction between the metal and semiconductor would not mimic the true properties of the hybrid. Nevertheless, such calculation would give some insight into the contribution from electromagnetic interaction to the final optical response of the HNS. In the T-matrix calculations, we used the experimentally measured sizes of the Ag NP and CdTe QD. In the case of Ag, the size of NP is sufficiently large enough to neglect the quantum confinement effects, and the bulk dielectric constants ($\varepsilon$) reported by Johnson and Christy\cite{Optical-constant-JohnsonChristy-prb-1972} were directly used for the T-matrix calculation. On the other hand, quantum confinement effect plays a vital role in modifying the dielectric constant of the CdTe QDs. Therefore to account for the quantum confinement effect, we estimate the modified dielectric constant of CdTe QD using the method reported by Marcelo Alves-Santos {\it et.al.} \cite{Dielectric-function-CdTe-trial-error-JPCC-2010}. This method uses a trial and error procedure to estimate the dielectric constants of CdTe QD using its bulk counterpart with the experimentally measured absorption spectrum. Figure.\ref{Fig:TmatrixAgnCdTe} shows the calculated extinction cross-section of a Ag NP surrounded by 32 CdTe QDs using T-matrix technique. For comparison, we have also shown the calculated extinction spectra of a single Ag NP in water and 32 non-interacting CdTe QDs. As expected, the optical response of Ag-nCdTe colloid is quite different with a strong red-shifted LSPR peak of about 75 nm compared to that of Ag NP colloid Further, the peak strength of LSPR peak for Ag NP also reduces to about 65\% of that of Ag NP. Such red-shift of LSPR peak and reduction in peak strength has been observed earlier in the hybrid formed by Ag nanoplate and CdTe QDs\cite{Counting_electron_Jayabalan_2019}. Thus, an excitation at 400 nm would only weakly excite the LSPR in case of Ag-nCdTe hybrid.

It is well established that the peak change in $\Delta T/T$ directly depends on the amount of energy absorbed by the metal NP\cite{Higher-nonlinearity-JOSAB-Jayabalan_2011,Counting_electron_Jayabalan_2019}. Any change in the amount of absorbed energy in presence of CdTe QDs should change the transient response of Ag NP in Ag-nCdTe hybrid. A comparison between the calculated extinction spectra of Ag NP and Ag-nCdTe hybrid can give an idea about the change in the absorbed energy when excited at 400 nm. The numerically calculated LSPR peak of Ag NP in water is at 393 nm while in the experiment, it is at 413 nm. Taking into account of such shift, the absorption in Ag NP in the hybrid should be only 4\% of that of bare Ag NP at 400 nm. The measured peak change in $\Delta T/T$ of Ag-nCdTe hybrid colloid is about 10\% of that of Ag NP colloid (Fig.\ref{Fig:400-408}). Further, an exponential decay fit to the measured $\Delta T/T$ in case of Ag-nCdTe gives 1.3 ps while it is 2.3 ps in case Ag NP colloid. The e-ph thermalization time depends on the absorbed energy in Ag NP. Thus a reduction in absorbed energy should also reduce the decay time of $\Delta T/T$ as observed in the case of Ag-nCdTe hybrid. Therefore, it looks tempting to attribute the observed changes in $\Delta T/T$ to the reduction in absorbed energy in Ag NP when surrounded by CdTe QDs in the hybrid colloid. 

\begin{figure}
	\includegraphics[width=0.5\textwidth]{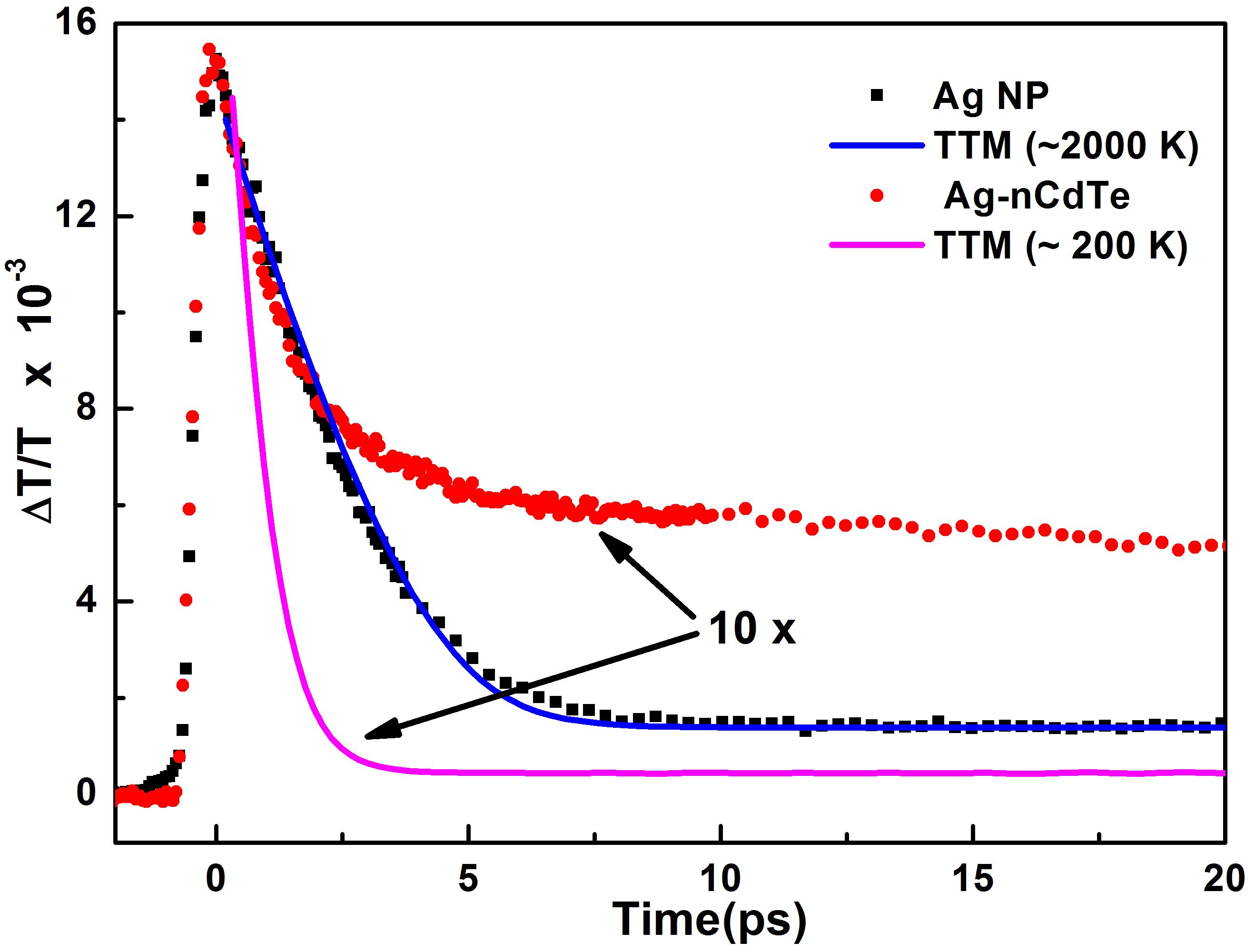}
	\caption{The TTM fit to the experimental $\Delta T/T$ measured for Ag NP colloid and Ag-nCdTe colloid. The data for Ag-nCdTe is increased by 10 times to make it comparable with the Ag NP.}
	\label{Fig:TTM-AgnCdTe}
\end{figure}

If the observed changes in $\Delta T/T$ is only due to such reduction in absorption as predicted by electromagnetic theory, the measured dynamics in Ag-nCdTe colloid can be explained using a two-temperature model (TTM), which works well for pure Ag NP colloid\cite{Higher-nonlinearity-JOSAB-Jayabalan_2011, Non-equilibrium-electron-dynamics-nobel-PRB-2000, Electron-dynamics-Chemical-Physics-2000}. Under low excitation conditions, change in dielectric constant of Ag NP is proportional to the temperature of the electrons\cite{Higher-nonlinearity-JOSAB-Jayabalan_2011}. The relation between the electron temperature (T$_{e}$) and lattice temperature (T$_{l}$) in TTM is given by following differential equations:
\begin{eqnarray}
\frac{\partial T_{e}}{\partial t} &=& -\frac{g}{C_{e}}\left( T_{e}-T_{l}\right) + \frac{Q(t)}{C_{e}} \label{Eq:TTM1} \\
\frac{\partial T_{l}}{\partial t} &=& \frac{g}{C_{1}}\left( T_{e}-T_{l}\right)
\label{Eq:TTM2}
\end{eqnarray}
where, $g$ represents e-ph coupling constant, $Q(t)$ is the absorbed power density, which is proportional to the pump intensity and $C_e$ (=$\gamma T_e$) and $C_l$ is the specific heat capacities of the free electrons and lattice respectively. Using the typical values reported for $g$, $C_l$, and $\gamma$ the Eq.\ref{Eq:TTM1} and Eq.\ref{Eq:TTM2} were solved numerically to obtain the time dependence of $T_e$. This estimated $T_e$ multiplied with a proportionality constant $C$ was then fit to the decay part of the measured $\Delta T/T$ with $Q$ and $C$ as fitting parameters. Figure.\ref{Fig:TTM-AgnCdTe} shows the best fit obtained for the decay part of the measured $\Delta T/T$ of Ag NP colloid. To fit the $\Delta T/T$ of Ag-nCdTe colloid, we keep all the parameters same as that used in the fitting of Ag NP colloid including the constant $C$ and reduce only $Q(t)$ until the peak $\Delta T/T$ matches that of the Ag-nCdTe colloid. The transient curve thus obtained is also shown in Fig.\ref{Fig:TTM-AgnCdTe}. Clearly, for the corresponding peak change, the fast decay time should have been much smaller than that observed in the experiment. Further, the magnitude of the long decay component for Ag-nCdTe should also be much lower than that observed in the experiment (inset of Fig.\ref{Fig:400-408}). Thus, the measured transient signal in Ag-nCdTe hybrid could not be just  attributed to the transient plasmonic response of Ag NP in the hybrid colloid. This suggests that the presence of CdTe QDs coupled with Ag NP in the Ag-nCdTe hybrid do play a role in dictating the final optical response. Several groups have measured the optical response of pure CdTe QDs \cite{Ultrafast-CdTe-PCCP-2010,Ultrafast-CdTeQD-JPCC-2010,Samanta-ultrafast-AgCdTe-JPCC-2016}. When CdTe QD colloid is pumped at 400 nm and probed on the red-side close to 400 nm, it shows a small increase in transmission (positive $\Delta T/T$). This increase in the transmission is attributed to the band-filling effect when carriers are excited into the conduction band of CdTe QD from its valance band \cite{Ultrafast-Absorption-II-VI-Nanowires-2009-JPCC,Ultrafast-CdTe-PCCP-2010}. Note that the bare CdTe QD colloid having same number density as that of Ag-nCdTe hybrid colloid, did not show any measurable signal. Nonetheless, an increased absorption in CdTe QD in presence of Ag NP near to it can contribute to the increased signal. Further, hot carriers from Ag NP can also get transported to CdTe QD in femtosecond time scale which can further increasing the transient response of CdTe QD in Ag-nCdTe hybrid. However, since all these proceseses contributes positively to the measured $\Delta T/T$, it is difficult to distinguish between the various contributions.

\begin{figure}
	\includegraphics[width=0.5\textwidth]{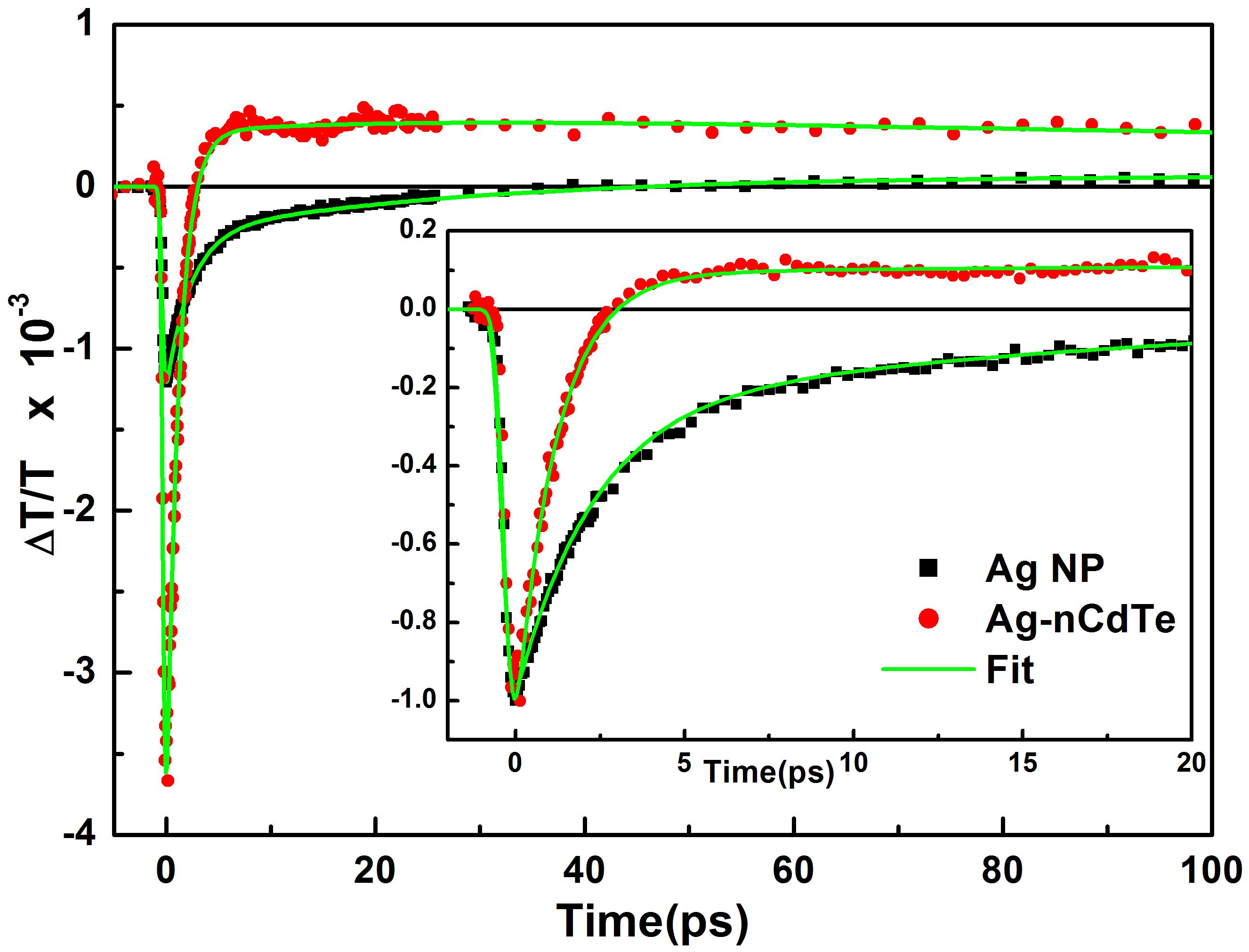}
	\caption{Transient transmission of Ag NPs and Ag-nCdTe hybrid colloidal samples when excited at 400 nm and probe at 550 nm. Inset shows the normalized curve of the same data.}
	\label{Fig:400-550}       
\end{figure}

When excited at 400 nm and probed close to the exciton peak, the CdTe QD colloid show a much larger positive $\Delta T/T$ signal when compared to that of probing near 400 nm\cite{Ultrafast-CdTe-PCCP-2010,Ultrafast-CdTeQD-JPCC-2010,Samanta-ultrafast-AgCdTe-JPCC-2016}. Further, when probed on the red-side of LSPR and well away from the LSPR,  Ag NP shows a weak negative $\Delta T/T$. Hence it should be possible to distinguish various processes if probed close to the 1s-1s exciton transition of the CdTe QD, i.e. at 550 nm. Figure.\ref{Fig:400-550} shows the transient transmission signal measured for the colloidal samples of bare Ag NP and Ag-nCdTe hybrid colloid when probed at 550 nm. The transmission of both the samples reduces with the arrival of the pump pulse ($\Delta T/T$  is negative). Once again a maximum of $|\Delta T/T|$ occurs by about 450 fs after the arrival of the pump pulse. Both the negative $\Delta T/T$ and the delayed maximum  are the signature of plasmonic contribution from Ag NP. In the next few tens of picoseconds, the $\Delta T/T$ of Ag NP colloid recovers with a time constant of 2.4 ps which is nearly same as that of e-ph thermalization time observed at 408 nm probing. On the other hand, in case of Ag-nCdTe colloid, the $\Delta T/T$ changes sign near about 3 ps, becoming positive at later times. This positive signal which could be a signature of the CdTe QD response, remains more or less constant up to 100 ps. The negative contribution of plasmonic response of Ag NP to the $\Delta T/T$ of Ag-nCdTe recovers within few picoseconds and is much faster than the positive contribution of CdTe QD. Thus at longer delays, the sign of $\Delta T/T$  becomes positive explaining the observed signal in $\Delta T/T$ at delays more than 3 ps in Fig.\ref{Fig:400-550}. To estimate the time constants the measured $\Delta T/T$ of Ag-nCdTe colloid at 550 nm was fit to the equation:
\begin{equation}
F(t) = \frac{1}{2} \mathcal{E}\left[ t/\tau_{r} \right)\left[- Ae^{ t/\tau_{1}} + B \left( 1-e^{ t/\tau_{1}} \right) + C \right],
\label{Eq:Fitting2}
\end{equation}
where, $\mathcal{E}$ is the error function. The first decay term corresponds to the plasmonic response of Ag NP and has a negative amplitude while the second term can be attributed to the positive contribution of band filling effect in CdTe QDs. We find that Eq.\ref{Eq:Fitting2} fits well to the measured $\Delta T/T$ with a time constant $\tau_{1}$ = 1.3 ps. This time constant is same as that of the decay constant of Ag-nCdTe hybrid when probed at 408 nm. Thus a model with enhanced absorption in CdTe QDs along with hot carrier dynamics in Ag NP can explain the observed changes in the measured transient $\Delta T/T$ at these probe wavelengths.

\begin{figure}
	\includegraphics[width=0.5\textwidth]{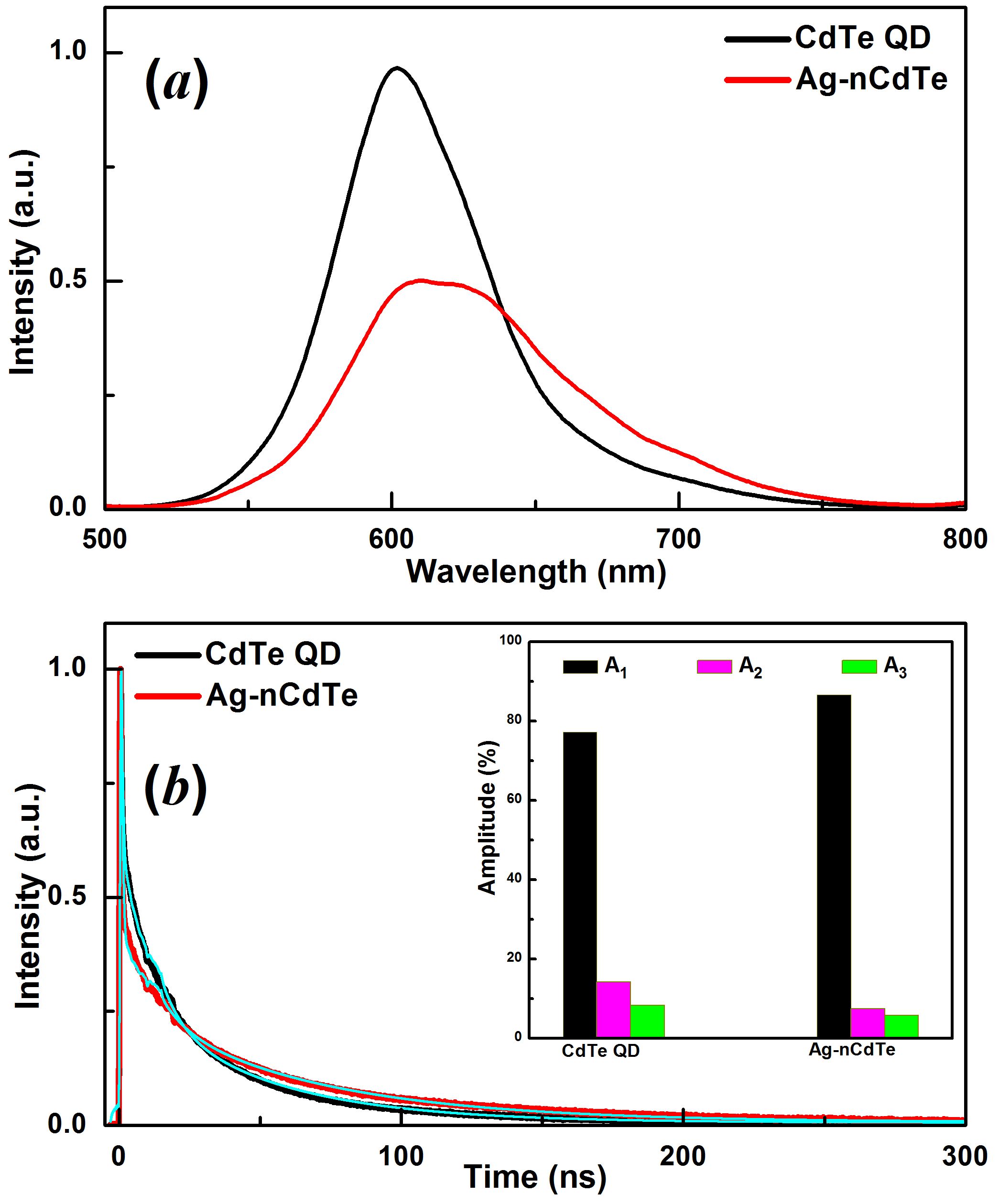}
	\caption{({\it a}) Photoluminescence spectra and ({\it b}) time-resolved photoluminescence of CdTe QD and Ag-nCdTe hybrid colloids. The inset shows the amplitudes of the best fit parameters obtained by fitting Eq.\ref{Eq:TriExpNSPL} to the experimental data.}
	\label{Fig:PL}
\end{figure}
Now, let us look at the effect of presence of Ag NP on the photoluminescence (PL) properties of the CdTe QDs. The PL measurement was carried out by exciting the sample at 405 nm. The PL emitted by the sample was collected using lens combination and detected by using a spectrograph(for spectrum measurement) or a fast photodetector-oscilloscope system(for dynamics). The photodetector has a rise time of 300 ps and the oscilloscope has a bandwidth of 1 GHz. The decay time estimated for the later part of the instrument response function (IRF) is 350 ps. If the field enhancement due to Ag NP increases the absorption in CdTe QD, after oscillation more number of electrons and holes should be present in the CdTe QD coupled to an Ag NP when compared to that of a bare CdTe QD. If all these electrons and holes recombine radiatively, there should be an increase in the PL emission from Ag-nCdTe hybrid compared to that of bare CdTe QD colloid. Figure.\ref{Fig:PL}({\it a}) shows the measured PL spectrum of CdTe QD colloid and Ag-nCdTe hybrid colloids. We find that the integrated area under the PL curve of Ag-nCdTe hybrid did not increase rather it got quenched to nearly 70\% of that of CdTe QD colloid. Figure.\ref{Fig:PL}({\it b}) shows the time-resolved PL measured for CdTe QD and Ag-nCdTe hybrid colloids. The time-resolved PL is fitted to tri-exponential function convoluted with the experimental IRF ($I_{IRF}$), given by
\begin{eqnarray}
F(t) = \left(I_{IRF} (\otimes) \sum_{i = 1,2,3}{A_{i} e^{\frac{-t}{\tau_{i}}} }\right)
\label{Eq:TriExpNSPL}
\end{eqnarray}
where A$_{i}$ is the signal amplitude corresponding to the decay time, $\tau_{i}$. Figure.\ref{Fig:PL}({\it b}) also shows the best fit obtained by fitting Eq.\ref{Eq:TriExpNSPL} to the experimental data. The fast decay time, $\tau_1$ = 0.23 ns, obtained for CdTe QD colloid  could be attributed to Auger like non-radiative relaxations while the other decay times 7 ns and 65 ns could be attributed to the raditive relaxation to the band-to-band recombination and defect state emission, respectively\cite{Auger_CdTe_PCCP_2013, Effect_Chloride_Passivation_CdTe_Binks_2015,Surface_related_emission_CdSe_Wang_2003,PL_Upconversion_CdTe_Xiao_2003}. The amplitude of each of these components are shown in inset of Fig.\ref{Fig:PL}({\it b}). The Ag-nCdTe hybrid also shows similar three component PL decay and the corresponding amplitudes are also shown in inset of Fig.\ref{Fig:PL}({\it b}). When compared with the CdTe QD colloid, the Ag-nCdTe hybrid shows an increase in the non-radiative component ($A_{1}$) while the radiative components $A_{2}$ and $A_{3}$ decreases. Although carrier density in CdTe QD is increased due to the presence of the Ag NP, such increase in carrier density can also increase Auger like recombinations\cite{Multiphoton_enhancement_CdSe_Au_film_Moyer_2013,SinglePhoton_emission_JPCC_2011}. In Auger recombination the electron and hole recombination energy is transferred non-radiatively to carriers deep into their corresponding bands. These deeply excited carriers then relax non-radiatively. Consequently, a fast quenching of radiative relaxation is observed in Ag-nCdTe hybrid colloid (Fig.\ref{Fig:PL}({\it a}).

\section{Conclusion}
A hybrid formation between a metal and semiconductor is known to significantly alter the static optical responses like absorption and PL spectra. We find that both non-radiative and radiative components of the individual entities also gets significantly altered when Ag NP and CdTe QD are brought in close contact. Transient transmission measurements performed by exciting at 400 nm and probing at two different wavelengths shows that the contribution of CdTe QD to the transient response got increased in presence of Ag NP.  However, such increase in the CdTe QD response did not result in increasing the radiative emission from the CdTe QDs. This work provides significant insight into the various relaxation processes that leads to the charge transport and PL quenching mechanisms in metal-semiconductor hybrids.

\section{Acknowledgment}
The authors are thankful to Tarun Kumar Sharma for fruitful suggestions and encouragement. The authors, Sabina Gurung and Durga Prasad Khatua, are thankful to RRCAT, Indore, under HBNI program, Mumbai, for the financial support.



\bibliographystyle{rsc} 
\bibliography{Article}

\end{document}